\documentclass[12pt,preprint]{aastex}

\shorttitle{Faint Fuzzies}
\shortauthors{Burkert, Brodie\& Larsen}

\begin{document}

\title{Faint Fuzzies and the Formation of Lenticular Galaxies}

\author{Andreas Burkert\altaffilmark{1}, Jean Brodie\altaffilmark{2} and Soeren Larsen\altaffilmark{3}}

\altaffiltext{1}{University Observatory Munich, Scheinerstrasse 1, D-81679 Munich, 
Germany}
\altaffiltext{2}{UC Observatories/Lick Observatory, University of California,
Santa Cruz, CA 95064, USA}
\altaffiltext{3}{European Southern Observatory, 85748 Garching, Germany}

\email{burkert@usm.uni-muenchen.de, brodie@ucolick.org, slarsen@eso.org}

\newcommand\msun{\rm M_{\odot}}
\newcommand\lsun{\rm L_{\odot}}
\newcommand\msunyr{\rm M_{\odot}\,yr^{-1}}
\newcommand\be{\begin{equation}}
\newcommand\en{\end{equation}}
\newcommand\cm{\rm cm}
\newcommand\kms{\rm{\, km \, s^{-1}}}
\newcommand\K{\rm K}
\newcommand\etal{{\rm et al}.\ }
\newcommand\sd{\partial}

\begin{abstract}

We investigate the dynamical state of a new class of extended star clusters, known as "Faint Fuzzies",
that were discovered in two nearby S0 galaxies, NGC~1023 and NGC~3384. It is shown that the
Faint Fuzzies of NGC~1023 lie in a fast rotating ring-like structure within the galactic disk
with mean radius of 5 kpc, rotational velocity of 200 km/s and velocity dispersion of 115 km/s.
We propose a scenario for the origin of Faint Fuzzies  that is connected to the origin of S0 galaxies 
as a result of galaxy-galaxy interactions in dense environments.
As is apparent in the Cartwheel galaxy, and is confirmed by numerical simulations, the passage of a small
galaxy through, or close to, the center of a disk galaxy can form a ring of clumpy star formation with a
radius comparable to the Faint Fuzzy ring radius in NGC~1023. In this case, the Faint Fuzzies are
signposts for the transformation of spiral galaxies into lenticulars via such interactions. 

\end{abstract}


\section{Introduction}

The discovery of the faint extended star clusters, now generally known
as the ``Faint Fuzzies'', hereafter FFs, has already been described in
some detail in \citet{lb00, bl02, bl03a, bl03b} along with detailed justifications of the assertion
that these objects are unique within the cluster "family". Here we provide only a brief summary of
their characteristics.

FFs were first discovered in an HST WFPC2 image of NGC~1023, a nearby (9.8 Mpc) SB0 galaxy \citep{lb00}.
They are distinct from normal compact globular and open clusters in a variety of respects. The effective half-light radius
R$_{eff}$ for the FFs is 7-15 pc while a typical R$_{eff}$ for globular and open clusters is $\sim$ 2-3 pc.
In addition to its FFs, NGC 1023 has a normal population of compact globular clusters which is
sub-divided into the red and blue sub-populations found in essentially all luminous galaxies \citep{lb02}.

\begin{figure}[!ht]
\plottwo{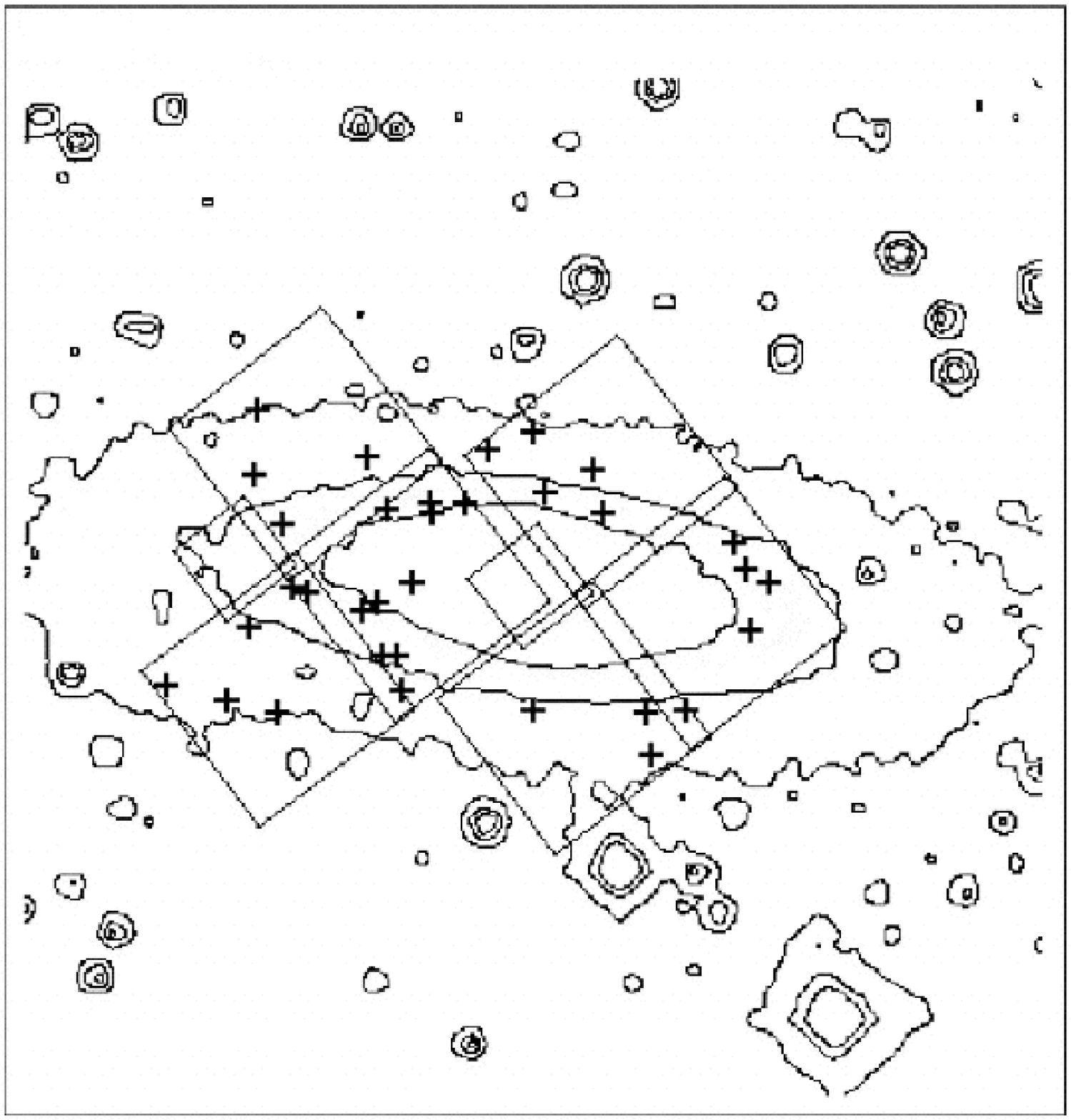}{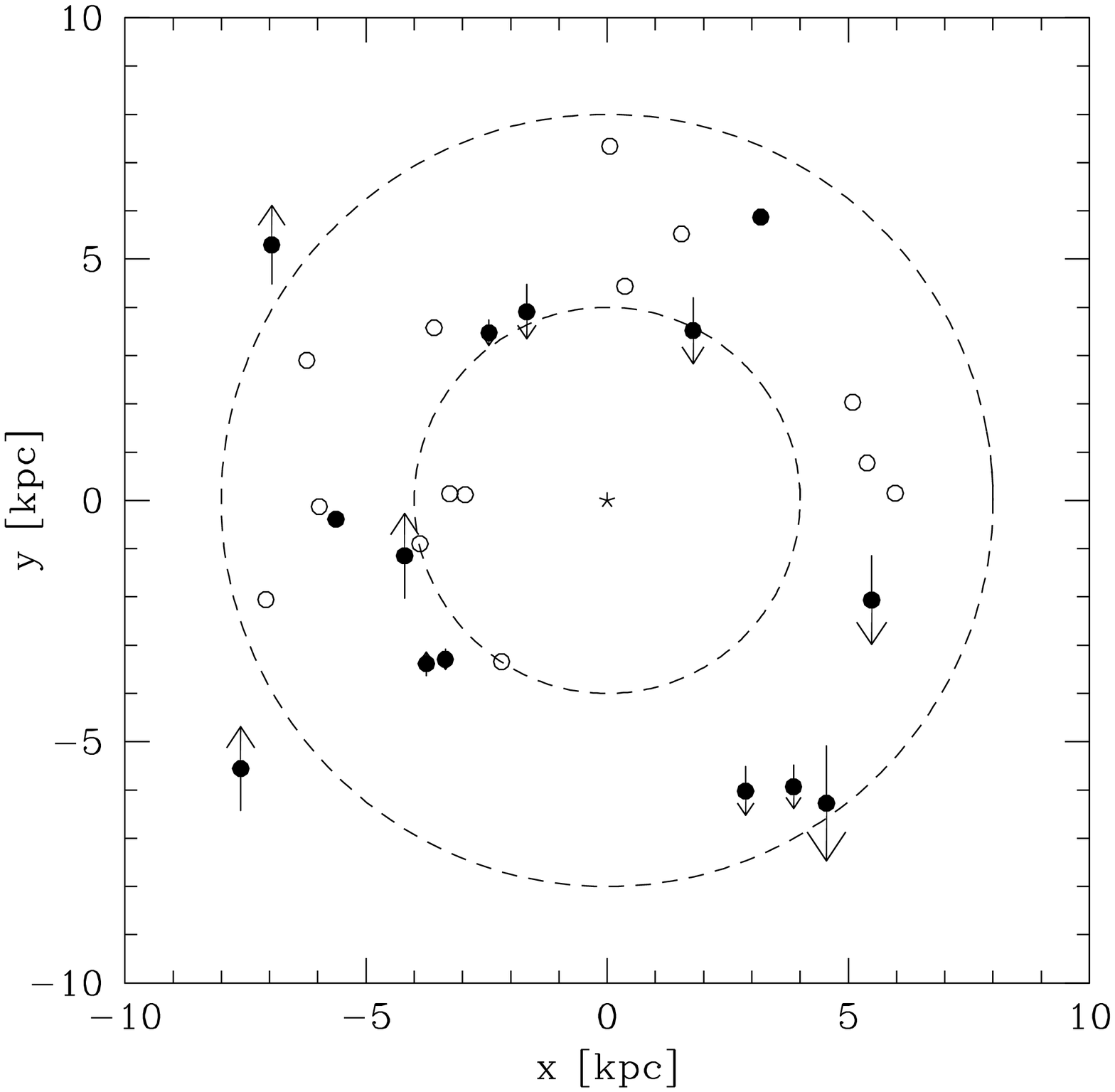}
\caption{
  \label{fig1}
Left panel: The contours of an optical image of NGC 1023 overlaid with the two WFPC2 pointings. The 
distribution of FFs (crosses) corresponds closely to the galaxy isophotes. Right panel: Deprojected location of
the FFs for a galaxy inclination of 66$^\circ$. Filled circles are FFs with measured line-of-sight velocities. Open circles are
FFs identified from the HST images but without measured velocities.  The arrows show the direction of 
rotation, inferred from the observed radial velocities, with the size of arrows being proportional 
to the measured line-of-sight velocity. }
\end{figure}

FFs have an annular distribution on the sky (see Fig.~\ref{fig1}), and are not nearly as concentrated toward
the center of the galaxy as the compact sources. It was noted in \cite{lb00} that they appeared to be
associated with the {\it disk} of NGC~1023, based on the high degree of alignment with
the galaxy isophotes. { \cite{lb00} also demonstrated that the lack of
FFs in the galactic center is not due to greater difficulties of detection near
the center. The annular distribution therefore seems to be real.}
NGC 1023 has 29 objects with half-light radii greater than 7 pc and
V$<$24 (i.e. M$_V$ brighter than --6.2, the limit for secure size
measurement). In \citet{l01}, a smaller number of faint, extended
objects similar to those in NGC~1023 were noted in another nearby
(11.5 Mpc) lenticular, NGC~3384.   An optical spectroscopic follow-up with the LRIS spectrograph \citep{o95} on
the Keck telescope confirmed the association of the FFs with both of these host galaxies, and allowed individual
radial velocites and a mean value of metallicity to be measured in a subset of the NGC~1023 and NGC~3384
samples \citep{bl02}.  They found [Fe/H] =$ -0.58\pm0.24$ and [Fe/H]=$-0.64\pm0.34$ for the FFs in NGC~1023
and NGC~3384, respectively.  The sample of 6 FFs observed in NGC~3384 is currently too small for a further
interpretation. { The few available data points indicate a ring-like
structure. However the velocity distribution does not show any signature of rotation.
More observations are required before any conclusion can be drawn about the kinematical and
spatial structure of the system of FFs in NGC 3384. In the following we therefore will concentrate
our analysis on the system of FFs in NGC~1023.}

An additional 7 candidate faint fuzzies were observed with Keck/LRIS in November 2004, using the
same instrumental set-up as in \citep{bl02}. The total exposure time was 7 hours, split into 1/2
hour integrations with a seeing of about 1 arcsec. Of the 7 objects 4 were too faint to obtain
usable spectra, 1 turned out to be a background galaxy at redshift $z=0.313$, thus leaving only
2 new FF radial velocity determinations. In addition, two of the FFs observed by \citep{bl02}
(N1023-FF-12 and N1023-FF-14) were re-observed during the Nov. 2004 run. The new measurements
for these clusters yield $539 \pm 21$ km/s and $676 \pm 13$ km/s, while the previous
measurements found $514 \pm 8$ km/s and $725 \pm 17$ km/s. Thus, the true velocity errors may be
somewhat underestimated, but no systematic differences are apparent between the old and new measurements.

\begin{figure}[!ht]
\plotone{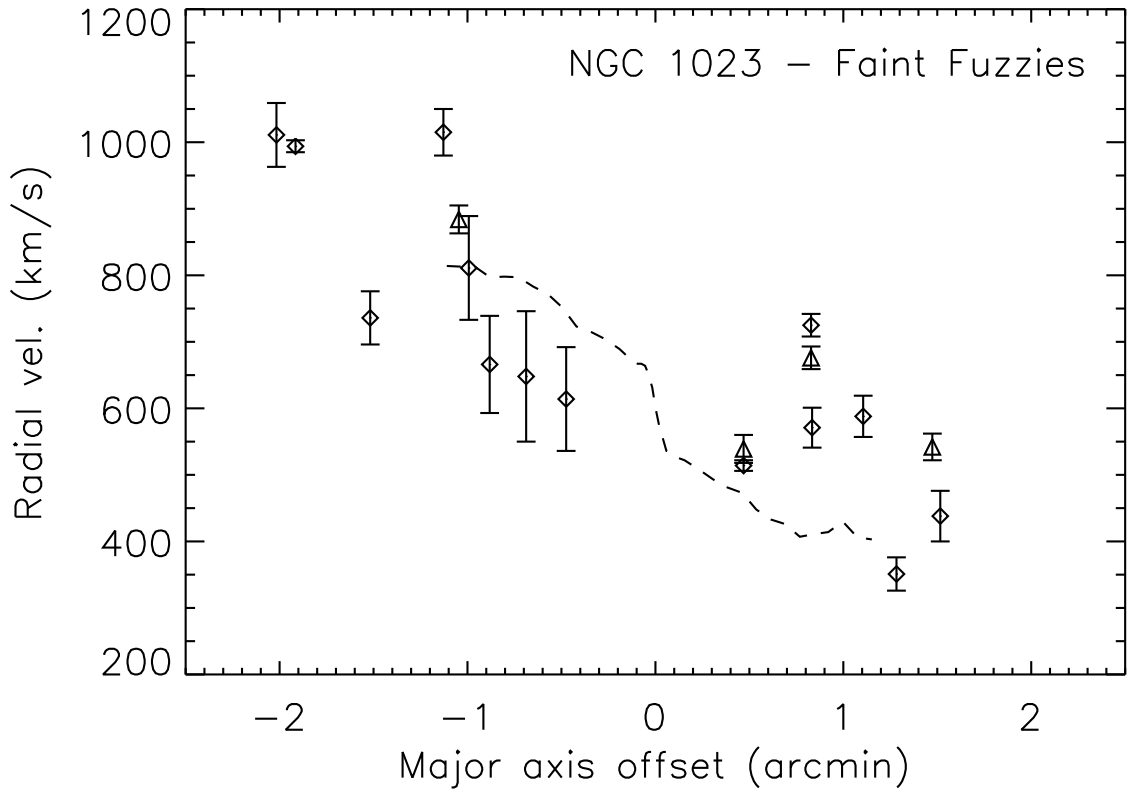}
\caption{
  \label{fig2}
Radial velocity vs.\ projected distance from
the galaxy center along the major axis for extended clusters in
NGC~1023. The triangles show the new measurements.
{ Note that the absolute values of the radial velocities differ
from those of Brodie \& Larsen (2002) who in a similar figure applied a correction of
-133 km/s to all data points, based on a comparison with some spectra taken with the 
red side on LRIS. This correction is omitted in this figure.} 
The dashed line indicates the rotation curve for NGC~1023
itself from a longslit positioned along the major axis \citet{sp97}.}
\end{figure}

Comparing measured values of [Fe/H] and H$\beta$ to stellar evolutionary models \citep{mt00},
\citet{bl02} found that the most probable age for these clusters is $\sim$13 Gyr, although the error on this
estimate was such that ages as low as 7-8 Gyr could not be ruled out. In any case, the FFs are clearly very old,
and thus highly stable against disruption,  which might indicate that they are on roughly circular orbits as this
would be likely to minimize disruptive tidal effects (disk/bulge
shocking). Indeed, Fig.~\ref{fig2}, the radial velocity plot for FFs in NGC~1023, shows clear evidence for
rotation of the cluster system with a kinematic signature that is similar but not identical to
the rotation curve of the galaxy itself (dashed curve), as measured along the 
major axis by \citet{sp97}. The bulge effective radius for 
NGC 1023 is $<$2 kpc \citep{mh01} so the clusters are $\sim$2 bulge effective
radii from the galaxy center ($1^\prime=2.9$ kpc at the distance of NGC~1023). This strongly suggests 
that these objects are associated with
the galaxy's disk rather than its bulge. In contrast to the FFs, there is no hint of rotation in the compact GCs in this galaxy.

While new HST ACS data can be expected to
reveal additional examples of FFs in galaxies out to Virgo distances, it is currently unknown
how common a phenomenon they represent, nor is it clear whether they are found exclusively in
S0s.  Of the 4 galaxies in the pre-ACS HST data archive with data quality sufficient for the detection of
FFs (deep images of nearby galaxies), FFs have been detected in two (NGC~1023 and NGC~3384) and ruled out
in the other 2 (NGC~3115 another lenticular) and in NGC~3379 (an elliptical).
If it turns out that FFs are found exclusively in lenticular galaxies
(a largely untested assumption) they might provide valuable insight into the formation
of this class of galaxies. In addition, clues about the origin (and/or survival) of FFs may perhaps
be found by asking what the SO galaxies NGC~1023 and NGC~3384 have in common
that differentiates them from the other lenticular NGC 3115. Interestingly, NGC~1023 is the dominant
member of a well-defined group of 15 galaxies and NGC~3384 is a member
of the Leo I group. By contrast, NGC~3115 is isolated except for a
dwarf companion at a projected distance of $\sim$17.5 kpc.
In addition, NGC~3115 is a highly bulge-dominated lenticular (actually
transitional between E7 and S01) whereas NGC~1023 and NGC~3384 are
both disk dominated SB0s.
  
\section{Kinematical analysis of the NGC~1023 sample}

\begin{figure}[!ht]
\plotone{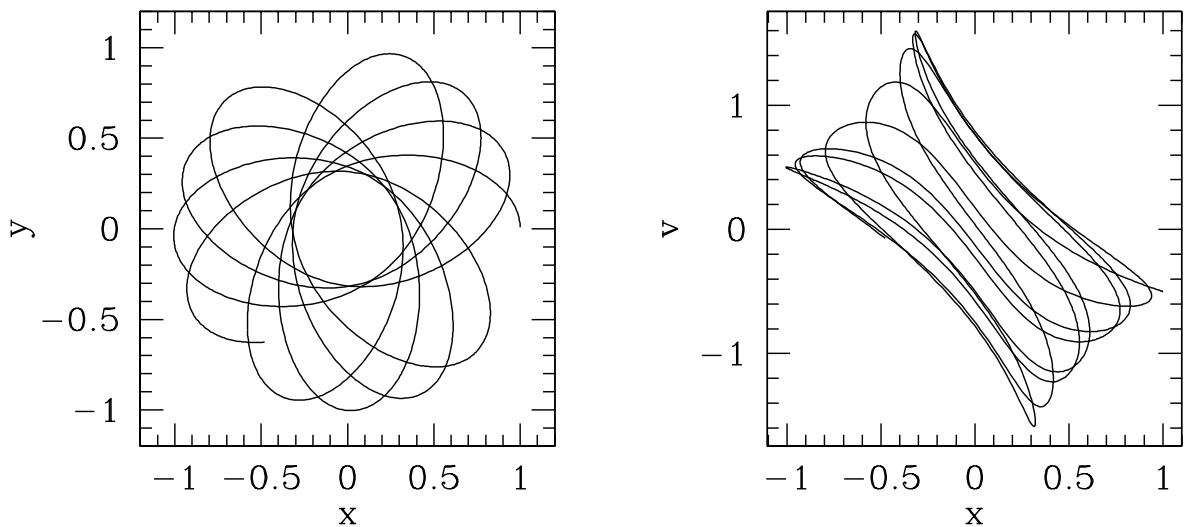}
\caption{
  \label{fig2orbit}
The left panel shows the typical elliptical orbit of a cluster in a logarithmic galactic potential. 
Its line-of-sight velocity as function of projected galactocentric distance is shown in the 
right panel.}
\end{figure}

The right panel of Fig. 1 and Fig. 2 show that the FFs of NGC~1023 rotate, however their rotation curve differs
significantly from that of NGC~1023. Instead of the expected extended region of constant rotation,
the average radial velocity changes linearly with major axis offset. This is the characteristic 
signature of a rotating ring,
seen in projection. In this case, the line-of-sight (here called radial) velocity 
$v$ would depend on projected major axis offset $x$ as

\begin{equation}
v(r) = v_{c} \times x/R + v_g
\end{equation}

\noindent where $v_g$ is the systemic velocity of the galaxy, $v_{c}$ is the deprojected rotational velocity of the 
ring and $R$ is its radius. A linear fit through the data points gives 
{ $v_g$=649 km/s $\pm$ 141 km/s which, within the uncertainties, is in good agreement with
the mean radial velocity of the galaxy (600 km/s) as determined by Simien \& Prugniel (1997).}
In addition, we find 
$v_{c}$  = 200 km/s and R = 5 kpc which agrees nicely with the rotational velocity of 
the galaxy at galactocentric distance of 5 kpc, as expected.  The system of FFs in addition has a significant 
dispersion in the line-of-sight velocities of $\sigma$ = 115 km/s, with some objects
even appearing to counter-rotate with respect to the rotational orientation of the galaxy. 

\begin{figure}[!ht]
\plotone{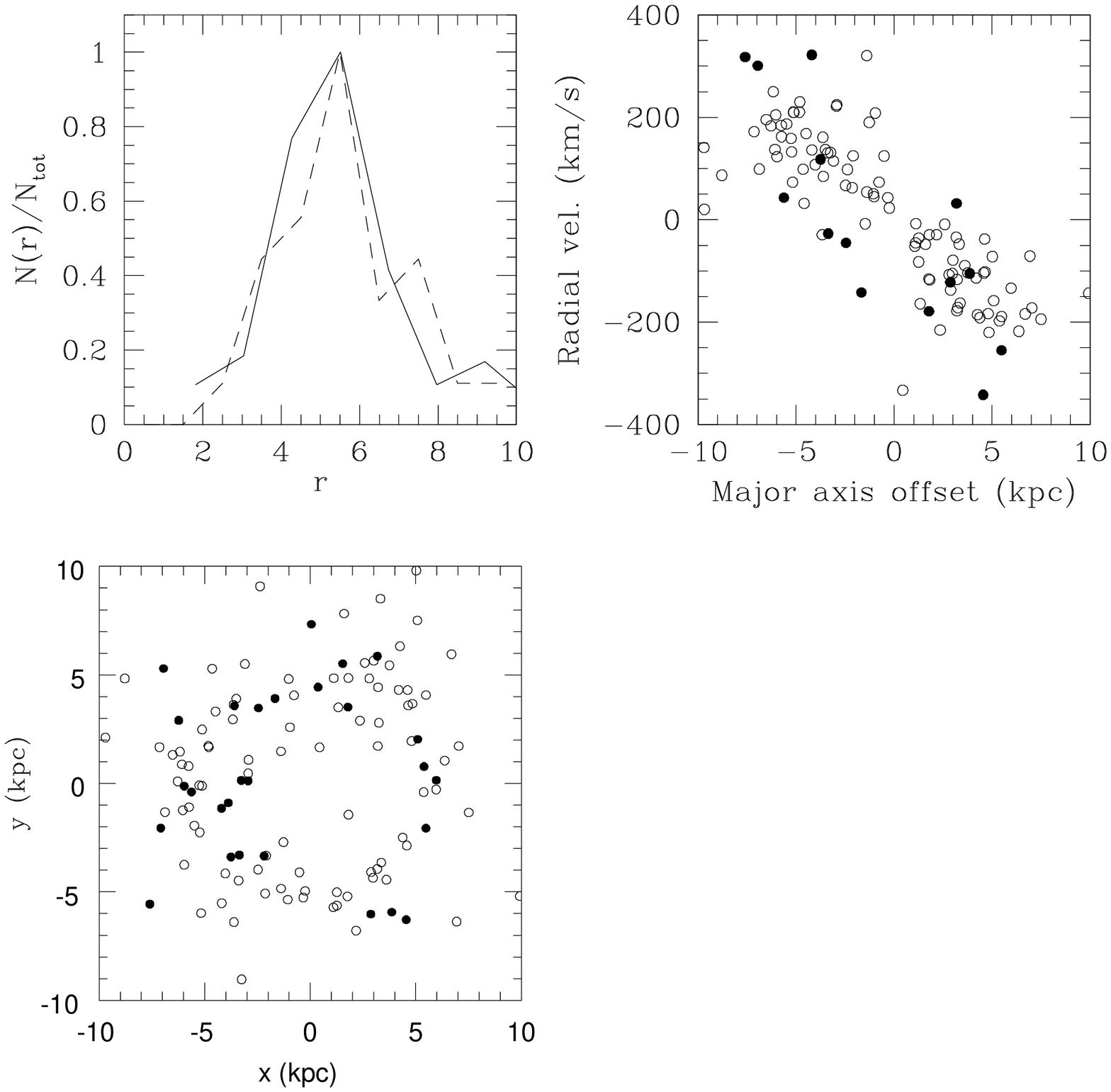}
\caption{
  \label{figkin}
The kinematical and spatial distribution of a phase-mixed stable system of point masses
forming in a ring with some initial velocity dispersion
is compared with the observed FFs of NGC 1023. The upper left panel compares the 
observed radial distribution (dashed curve) with the simulation (solid curve). The upper
right panel shows the distribution of line-of-sight velocities with filled circles
representing the observed sample and open circles showing the simulated point masses. The 
lower left panel shows the stable ring-like structure that has formed in the simulation 
and compares the spatial distribution in the disk of the simulated objects (open circles) with the 
observed FFs (filled circles).}
\end{figure}

This feature can be explained if the clusters move on elliptical orbits with mean radius $R$. 
As an example, figure 3 shows in the left panel the orbit of a cluster moving in a galactic 
logarithmic potential $\Phi(r) = v_{c}^2 \ln r$, where $r$ is the galactocentric distance.
Such a potential produces a constant rotation curve $v_{rot}(r) = v_c$. 
In the right planel the line-of-sight velocity of the cluster is plotted as
function of projected distance. Obviously, the width of the ring and the velocity dispersion are coupled.
In order to reproduce the observations we investigate a simple model where the clusters formed in a 
ring with radius R and width $\Delta R$. The clusters move in a logarithmic potential with $v_c=200 km/s$
and start initially within the ring with initial rotational velocity equal to $v_c$. An isotropic 
Gaussian velocity dispersion $\sigma$ is added to their rotational velocities. 
{ As a result of the additional velocity dispersion, the annular width of the
system will increase with respect to its initial width $\Delta R$. The final width of
the ring therefore is determined by the combination of the two free parameters, $\Delta R$ and
$\sigma$. If all objects formed at the same radius ($\Delta R = 0$), the width of the
ring would be determined only by the velocity dispersion. In the more realistic case of 
$\Delta R > 0$, the width of the ring is a combination of the initial width and the
velocity dispersion.}

The orbits are integrated until the cluster system is
phase mixed. Then its kinematical and spatial distribution is compared with the observations.
The upper left panel in figure 4 shows as dashed line the number distribution of the observed cluster system
as function of deprojected galactocentric distance $R=\sqrt{x^2+y^2}$, where x and y are the deprojected
disk coordinates of the clusters. The solid line shows the best fitting model which corresponds to
a ring-like distribution with radius R=5.3 kpc, width $\Delta R$ = 500 pc and
velocity dispersion of $\sigma$ = 80 km/s.
We find that the ring width $\Delta R$ mainly affects the width 
of the peak, whereas the shape of the wings are most sensitive to the adopted velocity dispersion.
Both a negligible ring width or a negligible velocity dispersion can be ruled out. The upper right and lower
left panels of figure 4 compare the line-of-sight velocity distribution and the deprojected location
of the modeled system of FFs (open points) with the observations (solid points). Note, that the ring-like
structure represents a stable solution which could in principle survive for many Gyrs if the disk is 
not perturbed, e.g. by a gravitational interaction with a satellite or another galaxy.

\section{Origin of the ring of clusters}

The kinematical and spatial distribution of the FFs in NGC 1023 indicates that they formed in the equatorial plane
in a fast-rotating ring-like configuration with width 500 pc and velocity dispersion of $\sim$80 km/s.
One possible solution for the origin of such a cluster ring is tidal disruption of FFs that are on orbits
with perigalactic distances smaller than 3 kpc. A good approximation of the rotation curve in the inner
3 kpc of NGC 1023 is

\begin{equation}
v_c(r) = 200 \left(\frac{r}{3 kpc}\right) km/s.
\end{equation}

\noindent Approximating the tidal radius $r_t$ of a cluster with mass M by (Binney \& Tremaine 1987)

\begin{equation}
r_t = r \left(\frac{M}{3 M_g(r)} \right)^{1/3},
\end{equation}

\noindent where 

\begin{equation}
M_g(r) =  \frac{v_c^2r}{G}
\end{equation}

\noindent is the total mass of the galaxy within radius r and G is the gravitational constant,
we find, that the tidal radius is independent of galactocentric distance r
and given by

\begin{equation}
r_t = 32 \left(\frac{M}{10^5 M_{\odot}} \right) pc.
\end{equation}

This value is large compard to the typical effective radii of the FFs (15 pc) which indicates
that tidal effects probably cannot explain the central hole. More detailed numerical simulations would
however be required to completely rule out this scenario.

The fact that the FFs
orbit in a ring-like configuration around the galaxy center is
difficult to explain if they arrived in ones and twos from parent
dwarfs originally on random incoming orbits. They are too
numerous for this to be a viable proposition and their metallicities
are too high for their origin in dwarf galaxies.
Nonetheless, it would be natural to suppose that more companion
galaxies were present in the past in the vicinity of NGC~1023 and
NGC~3384, so the presence of NGC~1023 and NGC~3384 in significant
groups may be relevant to a triggering mechanism for forming the
extended clusters {\it in situ} in the host galaxy. Taking this a step
further, the mechanism which led to the formation of lenticulars may
have induced faint fuzzy formation. 

There exists an intriguing similarity of the FF-ring with a similar feature in the 
Cartwheel galaxy that is believed to have formed through a central collision with another galaxy. 
The Cartwheel galaxy displays an inner ring of 
clumpy star formation with a radius ($\sim$4kpc) similar to the FF ring radius in NGC 1023.
Inspired by this observation \citep{am87} and numerical simulations \citep{hw93} of the development
of the star-forming rings and spokes in this galaxy, 
we speculate here that galaxy-galaxy interactions play a role in forming FFs.
In this case, group (or galaxy cluster) membership may be a relevant criterion for
selecting host galaxy candidates for FF searches. 
Note, that central encounters are rare. Off-centered encounters 
could in principle also trigger the formation lenticulars with FFs.
However the FFs would not organise themselves
into a stable ring-like distribution that lasts for several Gyrs.

\section{Discussion}
Stars and stellar clusters form in giant molecular clouds. FF masses are $\sim$10\% of
typical giant molecular cloud (GMC) masses. Their radii are
$\sim$10\% of a typical GMC radius. This suggests that FFs may have
formed inside a GMC. However, if every GMC produced a FF, they should
be very common.  In this case why are they not seen everywhere?  The
answer may be that, with only 10\% of the gas turning into stars, such star clusters
would remain bound only under certain (rare) conditions. Geyer \& Burkert (2002, 2005)
were able to create gravitationally
bound, long-lived star clusters with the sizes and masses of FFs,
provided that gas is compressed to densities of 
$n_{sf} \geq 10^3 cm^{-3}$ before star formation is allowed to start.
In this case, the resulting distribution of stars is very clumpy with
several sub-clusters. The sub-clusters are gravitationally bound and
merge into a virialized cluster with a typical mass $\sim 10^5 M_{\odot}$
and a typical radius of $\sim$10 pc. { \citet{fk02} presented a detailed
investigation of this process.  They noted that in interacting galaxies 
like the Antennae (e.g. Whitmore et al. 1999), young stars are found in small star clusters that are part
of larger groups (Harris 1998).  If these groups are gravitationally bound, 
successive mergers of their constituent subclusters would lead to new, larger
clusters which they called superclusters. Clearly, the superclusters are
larger than ordinary FFs which might be a result of the fact that the FFs
did not form in tidal arms. It is however possible that some central
triggering mechanism, maybe a central collision with another galaxy, generated an outwards moving
ring-like density enhancement which condensed into small star clusters.
This ring of cluster might 
later on have merge into larger objects. Numerical simulations, similar to
those presented by \citet{fk02} but for expanding ring-like structures would be required
to explore this scenario in greater detail.
Another interesting, unsolved problem is the difference between FFs and
normal, dense globular clusters. The origin of globular clusters is still not
well understood. Maybe globular cluster formation requires more violent trigger
mechanisms like supersonic cloud-cloud collisions which convert the gas of a GMC efficiently
into stars, instead of secular, dissipationless merging of small subclusters.}

The evolution of the morphology-density relation with redshift has recently been 
explored by \citet{s04} who find that the fraction of early type (E+S0) galaxies 
in dense environments has steadily increased from 70\% at z=1 to 90\% at the present epoch. 
Evolution in groups (regions of intermediate density) occurs only from z=0.5. No
evolution is seen in the field.  Smith et al.~attribute the evolution 
observed in intermediate and dense environments to the transformation of spirals into lenticulars.  
If FFs formed in this process they therefore should be substantially younger than those found in NGC 1023. If the gas fraction in the spirals was already
strongly reduced (e.g. by ram pressure stripping as a result of the galaxies' motion
through the intercluster medium) no FFs would have been able to form, although a central
encounter could still have converted the gas-poor spirals into S0s.

We have shown that the FFs of NGC 1023 are in a fast rotating, stable ring-like substructure
within the galactic disk which most likely was not formed by tidal disruption of objects
on highly eccentric orbits. In this case the most likely scenario is cluster formation in
a dense ring of metal-enriched gas, similar to young massive cluster formation in tidal
arms of interacting galaxies. More observations and numerical simulations are however
required to investigate the various ways that lead to FFs and to understand their 
possible connection to the origin of lenticulars in more details.  \\

\noindent
{ Acknowledgments:}
This work was supported by National
Science Foundation grant number AST-0206139.  A. Burkerts thanks the staff of the Astronomy
Department at the University of California, Santa Cruz for their warm hospitality during his
visits where part of this work was done.

\end{document}